\documentclass[preprint,notoc]{JHEP3}
\usepackage{graphicx}
\usepackage{amsmath}
\usepackage{epsf}
\usepackage{psfrag}
\usepackage[sort&compress,numbers]{natbib}

\def\be{\begin{equation}}
\def\ee{\end{equation}}
\def\bea{\begin{eqnarray}}
\def\eea{\end{eqnarray}}

\def\eqn#1{eq.~(\ref{#1})}

\def\Qb{{\bar{Q}}}

\def\Psl{\not{\hbox{\kern-2.3pt $P$}}}
\def\psl{\not{\hbox{\kern-2.3pt $p$}}}
\def\qsl{\not{\hbox{\kern-2.3pt $q$}}}
\def\Ksl{\not{\hbox{\kern-2.3pt $K$}}}
\def\ksl{\not{\hbox{\kern-2.3pt $k$}}}
\def\esl{\not{\hbox{\kern-2.3pt $\pol$}}}

\def\pol{\varepsilon}

\def\spa#1.#2{\left\langle#1\,#2\right\rangle}
\def\spb#1.#2{\left[#1\,#2\right]}
\def\lor#1.#2{\left(#1\,#2\right)}
\def\sand#1.#2.#3{%
\left\langle\smash{#1}{\vphantom1}^{-}\right|{#2}%
\left|\smash{#3}{\vphantom1}^{-}\right\rangle}
\def\sandp#1.#2.#3{%
\left\langle\smash{#1}{\vphantom1}^{-}\right|{#2}%
\left|\smash{#3}{\vphantom1}^{+}\right\rangle}
\def\sandpp#1.#2.#3{%
\left\langle\smash{#1}{\vphantom1}^{+}\right|{#2}%
\left|\smash{#3}{\vphantom1}^{+}\right\rangle}
\def\sandpm#1.#2.#3{%
\left\langle\smash{#1}{\vphantom1}^{+}\right|{#2}%
\left|\smash{#3}{\vphantom1}^{-}\right\rangle}
\def\sandmp#1.#2.#3{%
\left\langle\smash{#1}{\vphantom1}^{-}\right|{#2}%
\left|\smash{#3}{\vphantom1}^{+}\right\rangle}
\def\sandmm#1.#2.#3{%
\left\langle\smash{#1}{\vphantom1}^{-}\right|{\slash\!\!\! #2}%
\left|\smash{#3}{\vphantom1}^{-}\right\rangle}
\def\spab#1.#2.#3{\sandmm#1.#2.#3}
\def\spbb#1.#2.#3.#4{\sandpm#1.{\slash\!\!\! #2\slash\!\!\! #3}.#4}
\newbox\charbox
\newbox\slabox
\def\s#1{{      
        \setbox\charbox=\hbox{$#1$}
        \setbox\slabox=\hbox{$/$}
        \dimen\charbox=\ht\slabox
        \advance\dimen\charbox by -\dp\slabox
        \advance\dimen\charbox by -\ht\charbox
        \advance\dimen\charbox by \dp\charbox
        \divide\dimen\charbox by 2
        \raise-\dimen\charbox\hbox to \wd\charbox{\hss/\hss}
        \llap{$#1$}
}}
\def\ksl{\s{k}}

\def\beqa{\begin{eqnarray}}
\def\eeqa{\end{eqnarray}}
\def\beq{\begin{equation}}
\def\eeq{\end{equation}}
\def\hf{{\textstyle{1\over2}}}

\def\l{\lambda}

\def\vev#1{\langle{#1}\rangle}
\newcommand{\qn}[2]{\begin{equation} \label{#1} #2 \end{equation}}
\newcommand{\qalign}[2]{\begin{eqnarray}\label{#1} #2 \end{eqnarray}}
\newcommand{\Res}{{\rm Res}\,}
\newcommand{\wh}[1]{\widehat{#1}}
\def\CA{{\cal A}}

\def\CO{{\cal O}}

\def\l{\lambda}
\def\tl{\tilde\lambda}

\def\ket#1{|#1\rangle}
\def\MHVb{$\overline{\text{MHV}}$}

\def\BB#1#2#3{\;[#1|#2|#3\rangle}
\def\AA#1#2#3{\;\langle#1|#2|#3]}
\def\Aa#1#2#3{\;\langle#1|#2|#3\rangle}
\def\Bb#1#2#3{\;[#1|#2|#3]}
\def\A#1#2{\;\langle#1#2\rangle}
\def\B#1#2{\;[#1#2]}
\def\Qa{( (l_1+k_1)^2 - \mu^2 )}
\def\Qb{( (l_1+k_1+k_2)^2 - \mu^2 )}
\def\Qc#1{( (l_2+k_{#1})^2 - \mu^2 )}

\preprint{
  hep-th/0504159\\
  IPPP/05/13\\
  DCPT/05/26\\
  PUPT-2158\\
  April, 2005}

\title{Recursion Relations for Gauge Theory Amplitudes with Massive Particles}

\author{S. D. Badger,$^a$
    \ E. W. N. Glover,$^a$ V. V. Khoze${\,}^a$ and P. Svr\v{c}ek${}^b$\footnote{
    On Leave from Princeton University}\
    \\
        ${}^a$Department of Physics,
        University of Durham, Durham, DH1 3LE, UK\\
        ${}^b$Institute for Advanced Study, Princeton, NJ 08540, USA\\
        E-mails: \email{s.d.badger@durham.ac.uk,
    e.w.n.glover@durham.ac.uk,
        valya.khoze@durham.ac.uk, psvrcek@princeton.edu}.
    }

\abstract{We derive general
tree-level recursion relations for amplitudes
which include massive propagating particles. 
As an illustration, we apply these recursion relations
to scattering amplitudes of gluons coupled to massive scalars. We
provide new results for all amplitudes with a 
pair of scalars and $n\leq 4$ gluons.
These amplitudes can be used as building blocks in the computation of
one-loop 6-gluon amplitudes using unitarity based methods.}

\begin{document}

\section{Introduction}

The twistor string description of ${\cal N}=4$ Yang-Mills proposed
by Witten~\cite{Witten} has been an inspiration for much of the
recent progress in detailed calculations of multi-particle
amplitudes in gauge theory. New ideas have led to the development
of powerful new formalisms for these calculations, most notably
the MHV rules of Ref.~\cite{CSW1}, the Britto-Cachazo-Feng (BCF)
recursion relations of Refs.~\cite{BCFrec,Britto:2005fq}, and the
generalized unitarity cuts in the complexified Minkowski space of
Ref.~\cite{Britto:2004nc}. Applications of these new formalisms
together with the classic unitarity based approach of
Refs.~\cite{Bern:1994zx,Bern:1994cg} have led to a dramatic
progress in calculations of amplitudes.

At tree level, new and compact results for scattering amplitudes
were derived using the MHV
rules~\cite{CSW1,Zhu,GK,KosowerNMHV,BBK,GGK,Wu1,DGK,BFKM,BGK,Birthwright:2005ak,Roiban:2004ix},
and the recursion relations
\cite{BCFrec,Britto:2005fq,Luo:2005rx,Luo:2005my,Bedford:2005yy,Cachazo:2005ca,Britto:2005dg}.
External massive particles, such as Higgs bosons~\cite{DGK,BGK}
and electroweak bosons~\cite{BFKM}, have also been included using
generalizations of MHV rules. On the other hand, the BCF recursion
relations have been so far considered only for massless particles.

The first motivation of this paper is to construct tree-level
recursion relations which naturally incorporate massive particles.
The recursion relations follow from general quantum field theory
arguments, and are valid for any quantum field theory (and
also for gravity). The advantages of the recursion relations and the
MHV rules approaches over standard Feynman-diagram calculations
are obvious. Just as their massless parents
\cite{BCFrec,Britto:2005fq}, the recursion relations for massive
particles do not suffer from the factorial growth of the number of
contributing Feynman diagrams.
Compared to MHV rules, the recursion relations give more compact
results. They can be rigorously derived for a general quantum
field theory and they incorporate massive particles in a natural
and universal way. Essentially, with the general recursion
relations in place, one may be able to avoid in future all Feynman-diagrams
calculations of nontrivial tree amplitudes.

In this paper we will use massive recursion relations
to derive compact amplitudes for gluons coupled to massive coloured scalars. In the
forthcoming companion paper \cite{WIP} we will use the recursion
relations for amplitudes with massive particles with spin.

At one loop, the MHV rules have been successfully applied to
supersymmetric amplitudes
\cite{Cachazo:2004zb,Brandhuber:2004yw,Cachazo:2004by,Bena:2004xu,Cachazo:2004dr,
Quigley:2004pw,Bedford:2004py}.
New classes of amplitudes have been derived with the new methods
in
Refs.~\cite{Britto:2004nj,Bern:2004ky,Bidder:2004tx,Bidder:2004vx,Bedford:2004nh,
Britto:2004nc,
Bern:2004bt,Bern:2005hs,Bidder:2005ri,Bern:2005bb,Bjerrum-Bohr:2005xx,Britto:2005ha}.
All of these results give complete amplitudes only in
supersymmetric theories which are cut-constructible in 4 dimensions \cite{Bern:1994zx,Bern:1994cg}.
In non-supersymmetric gauge theories, the
new methods apply only to the 4D cut-constructible parts of the
amplitudes.

The second motivation of this paper is to assemble together the
pieces necessary for the complete calculation of one-loop
amplitudes in non-supersymmetric gauge theories.

One-loop multi-gluon amplitudes in the `all-plus', and in the `one-minus'
 helicity configurations,
$(+,+,+,\ldots,+)$ and $(-,+,+,\ldots,+)$, are known
\cite{Bern:1993qk,Mahlon:1993si,Bern:2005hs}.
However, the full one-loop
amplitudes even in the simplest non-supersymmetric gauge theory,
pure Yang-Mills, are still known only for $n\le 5$ external
gluons, with the five-gluon amplitude calculated in 1993 in
Ref.~\cite{Bern:1993mq}. The six-gluon amplitude has not yet been
calculated for all helicity configurations.
One-loop amplitudes with $n$ external gluons in
pure Yang-Mills are conveniently decomposed as
\be \label{decomp}
{\cal A}_n^{\rm gluon} = \, {\cal A}_n^{{\cal N}=4} - \, 4 {\cal
A}_n^{\rm chiral~{\cal N}=1} + \, {\cal A}_n^{\rm scalar} \ ,
\ee
where the first and the second terms are the contributions of the
${\cal N}=4$ supermultiplet, and the chiral ${\cal N}=1$
supermultiplet running in the loop. These contributions arise from
supersymmetric theories, they are cut-constructible and largely
known (at present all colour-ordered sub-amplitudes are known for
${\cal A}_n^{{\cal N}=4}$ for $n\le 7$,
\cite{Britto:2004nj,Bern:2004ky,Bern:2004bt},
and the complete ${\cal
A}_n^{\rm chiral~{\cal N}=1}$ is known for $n\le 6$,
\cite{Bern:1994zx,Bern:1994cg,Quigley:2004pw,Bedford:2004py,Bidder:2004tx,Bidder:2004vx,
Bidder:2005ri,Britto:2005ha}).

The last term in \eqref{decomp} corresponds to an $n$-gluon
amplitude with a complex scalar propagating in the loop. This is a
non-supersymmetric one-loop amplitude and it is not
cut-constructible \cite{Bern:1994cg}. What this means is that the
amplitude ${\cal A}_n^{\rm scalar}$ cannot be fully reconstructed
from its imaginary part evaluated in 4 dimensions. In other words,
the knowledge of the cuts of the amplitude is insufficient to
recover the full answer. The answer contains purely rational
terms, these do not have cuts in 4 dimensions.

To regulate infrared and ultraviolet divergent integrations over loop momenta,
loop amplitudes are commonly evaluated in $D=4-2 \varepsilon$ dimensions. In $D$ dimensions
(with non-integer $D$), the amplitudes contain {\em only}
cut-constructible contributions \cite{Bern:1995db,Bern:1lsdym}.
This is because all terms in the result are proportional to
$D$-dependent powers of the kinematic invariants, which give rise
to terms like $(-s)^{-\varepsilon}$ that necessarily contain
logarithms \cite{Bern:1995db}. Hence, in $D$-dimensions all
amplitudes are completely determined from their cuts and are
cut-constructible. The price to pay is that we now need to know
on-shell tree-level amplitudes in $D$-dimensions and work to
higher order in $\varepsilon$.

In order to determine full amplitude ${\cal A}_n^{\rm scalar}$ we
 cut the propagators in the loop and put them on-shell in
$D$-dimensions. On both sides of the cut we are left with
tree-level amplitudes of the form
\be \label{trscgl} {\cal
A}_{m+2}^{\rm tree}(\phi^\dagger_{l_1},g_1, g_2,
\ldots , g_m, \phi_{l_2} ) \ .
\ee
These are the colour-ordered
subamplitudes with two adjacent scalars and $m$ gluons of
arbitrary helicities. The external gluons have four-dimensional
momenta $p_1, \ldots, p_m$. The scalar momenta $l_1$ and $l_2$ are
$D$-dimensional, they are the loop momenta which we cut.

$D$-dimensional massless momenta of scalar fields can be thought
of as the $4$-dimensional on-shell momenta of particles with mass
$\mu$, such that $l_1^2=l_2^2=\mu^2$. Here $l_1$ and $l_2$ are
$4$-dimensional and the $\mu^2$ term arises from the extra
$D-4=-2\varepsilon$ dimensions \cite{Bern:1995db,Bern:1lsdym}.

Hence, in order to calculate 6-gluon one-loop amplitudes we need
to know tree-level $4,5,6$-point amplitudes \eqref{trscgl} with
$m=2,3,4$ gluons for all independent helicity arrangements. In
Ref.~\cite{Bern:1lsdym}, the amplitudes \eqref{trscgl} were
calculated for
up to $4$ gluons with the same helicity. 
Using the massive recursion relation we
will reproduce these results as well as derive the remaining
amplitudes \eqref{trscgl} with positive and negative helicity
gluons. \vskip .2 in \noindent{\it Notation}

We will be using the spinor helicity formalism
\cite{SpinorHelicity}. A vector is represented as a
bispinor
\qn{vector}
{p_{a\dot a}=\sigma_{a\dot a}^\mu p_\mu\ ,}
where $\sigma^\mu_{a\dot a}$ are the chiral gamma matrices. The
norm of $p$ is $p_\mu p^\mu= {\rm det}(p^{a\dot a}).$ Hence, a
vector is null if and only if it can be written as a product of
two spinors
\qn{massl}
{p^2=0 \ \Leftrightarrow \ p_{a\dot a}=\,
\lambda_a\tilde\lambda_{\dot a}\ .}
Here, $\lambda^a$ and
$\tilde\lambda^{\dot a},$ are commuting spinors of negative and
positive helicity respectively. For real momenta in Minkowski
signature $\tilde\lambda$ is a complex conjugate of $\lambda,$
hence $\lambda$ is conventionally called the ``holomorphic'' and
$\tilde\lambda$ is the ``anti-holomorphic'' spinor. When discussing
the recursion relations, it will be necessary to consider
complex momenta. For complex momenta, $\lambda^a$ and
$\tilde\lambda^{\dot a}$ are independent complex variables.

We will often use a short-hand notation:
\qn{nog}
{\lambda^a_i=\ket{i}^a\qquad \tl^{\dot a}_i=|i]^{\dot a}.}
The
invariant  spinor products are defined as
\be \label{Aspicon}
\vev{i~j} \equiv \ \vev{i^-|j^+} = \ \lambda_i^{~a} \lambda_{j~a}
 \ , \qquad
[i~j] \equiv \ \vev{i^+|j^-}  = \ \tilde\lambda_i^{~\dot a}
\tilde\lambda_{j ~\dot a} \ .
\ee
Here spinor indices are raised
and lowered with $\epsilon$-symbols.
The scalar product of a null vector $p_i$ with a vector $q$ is
\be
p_i \cdot q \, = \, - \hf \,\langle i|q|i] \ .
\ee

Throughout the paper we use the sign conventions\footnote{Note
that the opposite sign convention from \eqref{Aspicon} is used in
QCD literature for $[i~j]$.} of \cite{Witten,CSW1,BCFrec} and
define
\bea \label{1.4} s_{ij\ldots k} &=& (p_i+p_j+\ldots+p_k)^2
\\
\langle i | p |j] &=&  \langle i|^{a}\, p_{a}^{~\dot{a}} \,
|j]_{\dot{a}} = -  \lambda_i^{a}\, p_{a\dot{a}}\,
\tilde{\lambda}_j^{\dot{a}}
\\
\langle i | p_r p_s | j\rangle &=& \langle i|^{a}\,
p_{ra}^{~~\dot{a}} \,\, p_{s\dot{a}}^{~~b} \, |j {\rangle}_b =-
\lambda_i^a \, p_{r\,a\dot{a}}\,\,  p_{s}^{\,\dot{a}b}\,
\lambda_{j\,b}
\\ \label{1.7}
{} [ i | p_r p_s | j ] &=& [i|^{\dot{a}} \, p_{r\dot{a}}^{~~a}
\,\, p_{sa}^{~~\dot{b}} \, |j]_{\dot{b}} =-
\tilde{\lambda}_i^{\dot{a}} \, p_{r\,\dot{a}b}\,\,
p_{s}^{\,b\dot{b}}\, \tilde{\lambda}_{j\,\dot{b}} \eea For
massless momenta equations \eqref{1.4}-\eqref{1.7} become \bea
s_{ij} = \langle i\, j\rangle [i \, j] \ , \quad &&\langle i | p_r
|j] =  \langle i\, r\rangle [r\, j] \ ,
\\
\langle i | p_r p_s| j\rangle = \langle i\, r \rangle [r\, s]
\langle s\, j\rangle \ , \quad && [ i | p_r p_s | j ] =
  [ i\, r] \langle r\, s\rangle [s\, j] \ ,
\eea
and so on.

\section{Recursion relations with massive particles}
\label{sec:mrc}

Consider a tree level scattering amplitude of $n$ incoming
particles, some of which might be massive
\qn{amplitu}
{{\cal
A}(p_1,p_2,\dots, p_n)\ , \qquad p_i^2=m_i^2 \ .}
Single out two
particles,  $i,j$, for special treatment. These particles can be
either massive or massless. For given $p_i,p_j$ pick a null vector
$\eta=\lambda_\eta\tilde\lambda_\eta$ that is orthogonal to both
$p_i$ and $p_j$
\qn{conds}
{\eta\cdot p_i= \eta\cdot p_j=\eta^2=0\ .}
For generic $p_i$ and $p_j,$ there are exactly two such $\eta$
up to scaling. To see this, consider the plane spanned by $p_i$
and $p_j.$ Geometrically, the first two conditions in \eqn{conds}
mean that $\eta$ lies in the plane orthogonal to the plane spanned
by $p_i,p_j.$ The last condition sets $\eta$ to be in the
intersection of this plane with the lightcone. A generic plane
going through the origin intersects the complex lightcone at two
complex rays. These rays define the two solutions for $\eta$ up to
a scaling by a complex number. We now will construct these
solutions.

\vskip .2in \noindent{\it Solution for the shift momentum}

Let us find now explicit solutions of eqs.~\eqref{conds} for
complex momenta. We will discuss in turn the case when  both $i,j$
are massless, when one of $i,j$ is massive and at last the case
when both $i,j$ are massive.

If the marked momenta are null, $p_i=\lambda_i\tilde\lambda_i$,
$p_j=\lambda_j\tilde\lambda_j,$ then the condition that $\eta$ is
orthogonal to $p_i$ gives $2\eta\cdot p_i=\vev{\eta,i}[\eta,i]=0$,
so either $\lambda_\eta=\lambda_i$ or
$\tilde\lambda_\eta=\tilde\lambda_i,$. Similarly, vanishing of the
Lorentz invariant product of $\eta$ and $p_j$ implies
$\lambda_\eta=\lambda_j$ or $\tilde\lambda_\eta=\tilde\lambda_j.$
Combining these two conditions, we find two solutions
\qn{qnull}{
\eta=\lambda_j\tilde\lambda_i
 \ , \qquad
\eta'=\lambda_i\tilde\lambda_j \ .}

Now consider the case where the particle $i$ is massless and the particle
$j$ is massive. The condition that momentum $\eta$ is orthogonal to $p_i$
gives $\lambda_\eta=\lambda_i$ or
$\tilde\lambda_\eta=\tilde\lambda_i.$ If
 $\lambda_\eta=\lambda_i,$ the orthogonality to $p_j$ reads
\be
2\eta\cdot p_j=\, \lambda_i^{a} \, p_{ja\dot{a} }\,
\tilde\lambda_{\eta}^{\dot a} =\, 0\ ,
\ee
hence
$\tilde\lambda_\eta^{\dot a}=\lambda_{ia}\
p_j^{a\dot{a}}=(\lambda_i p_j)^{\dot a}.$ So the two possible null
vectors orthogonal to both $p_i$ and $p_j$ are
\qn{qtwo}
{\eta^{a\dot{a}}=\lambda_i^a(\lambda_{i}p_j)^{\dot a} \ ,
\qquad \eta'^{a\dot{a}}=
(p_j\tilde\lambda_{i})^a\tilde\lambda_i^{\dot{a}} \ .}
The case
when $i$ is massive and $j$ is massless is treated analogously.

The last case to consider is when both particles $i$ and $j$ are
massive. Here, neither $p_i$ nor $p_j$ is a product of two spinors
so the expression for $\eta$ is not as simple. We use the
condition $2\eta\cdot p_i=\lambda_\eta^a\,\tilde\lambda_\eta^{\dot
a}\, p_{ia\dot a}=0$ to express $\lambda_{\eta a}=p_{ia\dot a}\,
\tilde{\lambda}_\eta^{\dot a}.$ Putting this into the second
orthogonality condition gives a quadratic equation for
$\lambda_\eta$
\qn{quq}
{ \lambda_\eta^a \, \lambda_\eta^b \,
p_{ia\dot a}\, p_{jb\dot b}\, \epsilon^{\dot a\dot b}=\, -\langle
\eta|p_i\cdot p_j|\eta\rangle = 0\ .}
This equation has two
solutions, $\lambda_\eta^{\pm},$ which we can find, for example,
by setting $\lambda_\eta^a=(1,x)$ and solving the quadratic
equation for $x.$ The analogous condition for the positive
helicity spinor $\tilde\lambda^{\dot a}_{\eta}$
\qn{neq}{
\tilde\lambda_\eta^{\dot a}\, \tilde\lambda_\eta^{\dot b}\,
p_{ia\dot a}\, p_{jb\dot b}\, \epsilon^{ab}=\, -[\eta|p_i\cdot
p_j|\eta]=\, 0}
has also two solutions. Altogether, up to scaling,
there are two null vectors $\eta=\lambda_\eta\tilde\lambda_\eta$
that are orthogonal to $p_i,p_j.$ We do not know of a convenient
Lorentz invariant solution to \eqn{quq} and
\eqn{neq}. This makes the case where both marked particles are
massive less tractable than the two simpler cases where at
least one of the marked particles is lightlike.

\subsection{Derivation of the recursion relations}
\label{sec:derrc}

To construct massive recursion relation for a tree-level $n$-particle amplitude
${\cal A}(p_1,p_2,\dots, p_n),$ we first mark particles $i$ and $j$ for special treatment
and pick one of the two null
vectors $\eta$ satisfying the conditions in eqs.~\eqref{conds}.
Following  ~\cite{Britto:2005fq}, consider the auxiliary
function of one complex variable
\qn{aux}{{\cal A}(z)={\cal A}
(p_1(z),\dots, p_i(z),\dots, p_j(z),\dots, p_n(z))\ ,}
where
$p_k(z)=p_k$ for $k\neq i,j$, and
\qn{pzet}{p_i(z)=p_i+z\eta \ ,
\qquad p_j(z)=p_j-z\eta\ .}
Since $\eta$ is null and orthogonal to
$p_i,p_j$, the shifted momenta are on-shell
\qn{onshl}{p_i(z)^2=p_i^2\ ,\qquad p_j(z)^2=p_j^2\ .}
Equations~(\ref{pzet}) imply that
$p_i(z)+p_j(z)=p_i+p_j,$ so ${\cal A}(z)$ obeys momentum
conservation. Hence, it is an on-shell scattering amplitude of
particles with complex momenta and can be computed from the usual
Feynman rules.

Clearly, the momenta of the external particles are linear
functions of $z.$ Notice that the spinors of massless external
particles are linear functions of $z$ as well. In the case where
both marked particles are massless, there are two possible
$\eta$'s given by eq.~\eqref{qnull}. For
$\eta=\lambda_j\tilde\lambda_i,$ the shift \eqref{pzet} is
accomplished by \qn{lzet} {\lambda_i(z)=\lambda_i+z\lambda_j \ ,
\qquad \tilde\lambda_j(z)=\tilde\lambda_j-z\tilde\lambda_i \ .}
The second solution for the shift vector,
$\eta'=\lambda_i\tilde\lambda_j,$ gives
\qn{eq210b}{\tilde\lambda_i(z)=\tilde\lambda_i+z\tilde\lambda_j \
, \qquad \lambda_j(z)=\lambda_j-z\lambda_i \ .}

Consider now the case when one of the particles, say particle $i$,
is massless and the other particle $j$ is massive. Then \eqn{qtwo}
gives $\eta^{a\dot{a}}=\lambda_i^a(\lambda_{i}p_j)^{\dot a}.$ The
shift of marked momenta \eqref{pzet} is accomplished by
\qn{qtwos}{\tilde\lambda_i^{\dot a}(z)=\tilde\lambda_i^{\dot a}
+z(\lambda_{i}p_j)^{\dot a} \ , \qquad
p_j^{a\dot{a}}(z)=p_j^{a\dot{a}}-z
\lambda_i^a(\lambda_{i}p_j)^{\dot a}\ .}
For $\eta^{a\dot{a}}=
(p_j\tilde\lambda_{i})^a\tilde\lambda_i^{\dot a}$ there are
analogous expressions
\qn{qtwosa}{\lambda_i^a(z)=\lambda_i^a
+z(p_j\tilde\lambda_{i})^a \ , \qquad
p_j^{a\dot{a}}(z)=p_j^{a\dot{a}}-z
(p_j\tilde\lambda_{i})^a\tilde\lambda_i^{\dot a}\ .}

It follows that ${\cal A}(z)$ is a rational function of $z$
because at tree level, the scattering amplitude is a rational
function of the spinors of massless external particles and of the momenta
of massive external particles.

\begin{figure}[h]
    \psfrag{i}{$i$}
    \psfrag{j}{$j$}
    \psfrag{r}{$r$}
    \psfrag{rm1}{$r-1$}
    \psfrag{s}{$s$}
    \psfrag{sp1}{$s+1$}
    \psfrag{P}{$P$}
    \begin{center}
        \includegraphics[width=8cm]{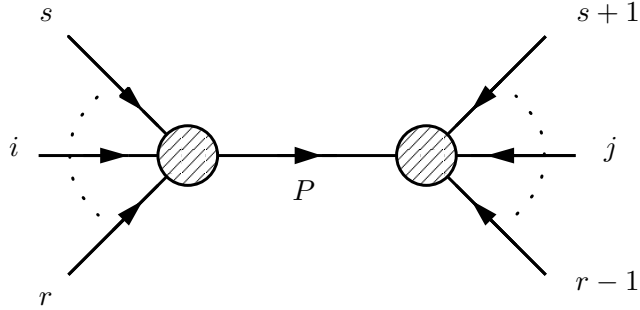}
    \end{center}
    \caption{Diagrammatic representation of the recursion relation. Arrows label the
    momentum flow.}
    \label{fig:recurs}
\end{figure}

At tree-level the rational function ${\cal A}(z)$ can only have simple poles in $z$
coming from internal propagators $1/P(z)^2$. Each propagator
divides the external particles into two groups, the particles to
the `left' and to the `right' of the propagator as illustrated in
figure \ref{fig:recurs}. Hence, the momentum $P(z)$ of a
propagator is the sum of the  momenta of the external particles to
the left of the propagator
\qn{summ}
{P=p_{r}+\ldots+p_i+\ldots +p_{s}\ .}
Momentum $P(z)$ depends on $z$ only if the particles $i$
and $j$ are on opposite sides of the propagator. We choose the
particle $i$ to be on the left of the propagator and the particle
$j$ to be on the right, as in figure \ref{fig:recurs}. Then \be
P(z)=P+ z\eta \ , \ee and the propagator is, \qn{polp} {{1\over
P(z)^2-m^2} \, =\, {1\over P^2-m^2+2z P\cdot \eta}\ ,} where $m$
is the mass of the internal particle. The propagator \eqref{polp}
has a simple pole at \qn{pole} {z\, =\, -\, {P^2-m^2\over 2 P\cdot
\eta} \ .} For generic external momenta, all internal momenta are
different, hence, the locations of all poles are different. It
follows that the tree-level amplitude ${\cal A}(z)$ has only
simple poles as a function of $z.$

To find the recursion relations, we use the familiar theorem from
complex analysis that the sum of residues of a rational function
on a Riemann sphere is zero. Applying this to ${\cal A}(z)/z$ we
express ${\cal A}(0)$ as a sum over residues
\qn{ao}
{{\cal
A}(0)=\, \Res \left({{\cal A}(z)\over z}\right)_{z=0}=\, -\sum_\alpha
\Res \left({{\cal A}(z)\over z}\right)_{z=z_\alpha}- \Res
\left({{\cal A}(z)\over z}\right)_{z=\infty}\ ,}
where the sum is
over all finite poles $z_\alpha$ of the amplitude ${\cal A}(z)$.
These come from the propagators $1/P^2(z)$ that separate the
particles $i$ and $j.$ The residues at finite $z$ are determined
by the factorization of the scattering amplitude when the Feynman
propagator \eqref{polp} goes on-shell
\qn{finz}
{\Res \left({{\cal A}(z)\over z}\right)_{z=z_\alpha} = \,
-\,{{\cal A}_L(z_\alpha)
{\cal A}_R(z_\alpha)\over P^2-m^2}\ .}
Here, ${\cal A}_L$ and ${\cal
A}_R$ are the tree-level amplitudes of the particles to the left
and to the right of the propagator and $z_\alpha$ is given by \eqn{pole}.

Hence, any tree-level scattering scattering amplitude ${\cal
A}={\cal A}(0)$ can be written in the form
\qn{expr} {{\cal
A}=+\sum_\alpha {{\cal A}_L(z_\alpha) {\cal A}_R(z_\alpha)\over
P_\alpha^2-m_\alpha^2}-\Res\left({{\cal A}(z)\over
z}\right)_\infty\ ,}
where the sum is over all channels $\alpha$
such that the particles $i$ and $j$ are on different sides of the
channel, and $z_\alpha$ is given by \eqn{pole}. The relation
\eqref{expr} is useful for computing scattering amplitudes only if
there is an efficient way to determine the boundary contribution
$\Res({\cal A}(z)/z)_\infty.$ The most favourable scenario is when
this contribution vanishes. This happens if and only if ${\cal
A}(z)$ vanishes at infinity,
\qn{rif} {\Res \left({{\cal
A}(z)\over z}\right)_{\infty}=0 \quad \Leftrightarrow \quad {\cal
A}(z)\rightarrow 0,\,\,{\rm for} \,\, z\rightarrow \infty\ ,}
in which case, there is a simple recursion relation:
\qn{recrel}
{{\cal A}=\sum_\alpha {\cal A}_L(z_\alpha) {1\over
P_\alpha^2-m_\alpha^2} {\cal A}_R(z_\alpha)\ }
that expresses
$\CA$ in terms of lower-point on-shell scattering amplitudes
$\CA_L(z_\alpha)$ and $\CA_R(z_\alpha).$  The summation in
\eqref{recrel} runs over all partitions of particles between
${\cal A}_L(z_\alpha)$ and ${\cal A}_R(z_\alpha)$, such that $p_i$
is on the left, and $p_j$ is on the right of $P_\alpha$, and also
over all helicities $h$ of the intermediate state $P_\alpha$.

The above considerations apply to the case with massive or massless marked
particles. However, for calculations carried out in this paper it
will be sufficient to take both marked particles to be massless.
In this case, the necessary conditions  for the vanishing of the
boundary contribution \eqref{rif} put constraints on the possible
helicities of the marked particles $i$ and $j$. We discuss these
conditions in section {\bf \ref{vanish}}.

\subsection{Recursion relations: summary}
\label{sec:sumrc}

We will use the recursion relations to calculate tree-level
scattering amplitudes in Yang-Mills theory coupled to matter
fields. The matter fields may be massive or massless and transform
in a generic representation of the gauge group. We consider the
colour-ordered partial amplitudes ${\cal A}={\cal A} (p_1,\dots,
p_n),$ in which the coloured particles come in a definite cyclic
order $1,2,\dots, n$. These amplitudes are obtained by stripping
away the colour factors from the full amplitude. hence, they depend
on the kinematic variables, momenta and helicities, $p_k$ and $h_k$ only.

In the remainder of the paper we will take both marked particles
to be massless. We shift two massless momenta ${p}_i=\ket{{i}}
|{i}]$ and ${p}_j=\ket{{j}}|{j}]$ of the marked particles by
 $\eta=\ket{j}|i]$, so the shifted momenta are
\bea \label{eq:225}
 \widehat{p}_i &=&  p_i + z |j\rangle|i]\ , \\
\label{eq:226}
\widehat{p}_j &=&  p_j - z |j\rangle|i] \ , \\
\label{eq:227} \widehat{P} &=& P + z |j\rangle|i] \ , \eea where
$P = p_r + \ldots + p_i +\ldots + p_s$ is the momentum of the
intermediate particle. For the particles $i,j$ this is equivalent
to shifting the spinors
\begin{eqnarray}
&&|\widehat{i}] = |i]  \ , \qquad \, \, |\widehat{i}\rangle = |i\rangle
+ z |j\rangle \ ,
    \label{eq:shiftsi} \\
    &&|\widehat{j}\rangle = |j\rangle \ , \qquad
    |\widehat{j}] = |j]- z |i] \ .
    \label{eq:shiftsj}
\end{eqnarray}
The recursion relation \eqref{recrel} written more explicitly is
(c.f. figure \ref{fig:recurs})
\bea
    \mathcal{A}_n (p_1,\ldots,p_n) =\,
    \sum_{{\rm partitions}}\sum_{h}
    && \mathcal{A}_L(p_r,\ldots,\wh{p}_i,\ldots,p_s,-\wh{P}^h) \,
\frac{1}{P^2-m_p^2} \nonumber \\
   &&\times \mathcal{A}_R(\wh{P}^{-h},p_{s+1},\ldots,\wh{p}_j,\ldots,p_{r-1}) \ ,
    \label{eq:rrel}
\eea
where summation is over all partitions of $n$ external particles
between ${\cal A}_L$ and ${\cal A}_R$, such that $p_i$ is on the
left, and $p_j$ is on the right, and also over the helicities,
$h,$ of the intermediate state.  $z$ can be found from the
on-shell condition $\widehat{P}^2=m_p^2$,
 \be
 \label{eq:228}
 z =\,  -\, {{P}^2-m_p^2\over 2 {P} \cdot \eta} =\,
\frac{P^2-m_p^2}{ \AA{j}{P}{i} } \ .
\ee

When the intermediate state $P$ is a massless particle (e.g. a
gluon), we can simplify the spinor products involving
$|\widehat{P}\rangle$ and $|\widehat{P}]$ as in
Ref.~\cite{BCFrec}:
\bea \label{eq:231}
 \langle k~\widehat{P} \rangle &=&
 \frac{\langle k | \widehat{P} |i]}{[\widehat{P}~i]} \equiv \,
\frac{\langle k | P | i]}{\omega} \ ,\\
\label{eq:232} {}[\widehat{P} ~ k] &=& {} \frac{\langle j |
\widehat{P} | k]}{\langle j ~\widehat{P}\rangle}
 \equiv \,
\frac{\langle j | P | k]}{\overline{\omega}}\ ,
\eea
where $\omega$
and $\overline{\omega}$ enter the amplitude always in the
combination $\omega \overline{\omega} = \langle j | P | i]$.

For practical computations it is essential that ${\cal A}(z)$
vanishes for large $z$, so that the recursion relations do not
have a boundary contribution at infinity. As discussed in section
{\bf \ref{vanish}}, this puts a constraint on the helicities of
the particles  $i$ and $j$.
For
our choice of the shift momentum, $\eta=\ket{j}|i],$ the helicities of the
marked particles can take the values,
\be \label{hconds}
\eta=\ket{j}|i] \, : \qquad (h_i,h_j)\,
=\,(+,-)\ ,\ (+,+)\ , \ (-,-)
\ee
but not $(h_i,h_j)=(-,+).$
Conditions \eqref{hconds} are the same for massive and for massless amplitudes.


\section{Amplitudes with gluons and massive scalars}
\label{sec:four}

In this section we consider scattering amplitudes of gluons with
massive complex scalars. These amplitudes are related to
amplitudes with {\it massless} scalars that have $D$-dimensional
momenta. The scalars with $D$-dimensional momenta $P_{D}$ can be
thought of as massive scalars in 4 dimensions. The $D$-dimensional
on-shell condition, ${P_{D}}^2=0$, gives the $4$-dimensional
mass-shell equation, ${P_{4}}^2=\mu^2$, where the mass term,
$\mu^2$, arises from the extra $D-4$ dimensions of momenta.

We will derive  amplitudes with 2 scalars and up to 4 gluons with
arbitrary helicity configurations. The amplitudes
with the same-helicity gluons have been previously derived in
\cite{Bern:1lsdym}.

\subsection{Primitive vertices}

The recursion relations construct $n$-point amplitudes from
on-shell $m$-point amplitudes with $m<n$. The $m$-point amplitudes
are connected to each other with scalar propagators. Using the
recursion relation $n-3$ times gives a representation of the
$n$-point amplitude entirely in terms of the 3-point vertices.
Hence, $3$-point vertices are the building blocks of the
amplitudes in the recursive approach\footnote{In particular, this
implies that the 4-point vertices in the microscopic Lagrangian
are not used in the recursive construction of gauge-invariant
amplitudes \cite{BCFrec}.}, they will be called the primitive
vertices.

The recursion relation reduce the task of computing general
amplitudes to the computation of all 3-point primitive vertices.
In this paper we consider amplitudes with massless gluons $g$ and
massive scalars $\phi$. These can be built from the  $ggg$ and
$\phi g \phi^\dagger$ vertices.

The three-gluon primitive amplitudes can have $--+$ or $++-$
helicity configurations. These are the standard MHV and
$\overline{\rm MHV}$ 3-point on-shell amplitudes \be
\mathcal{A}_3(g_1^-,g_2^{-},g_3^+) = \,
\frac{{\A{1}{2}}^3}{\A{2}{3}\A{3}{1}}\ , \qquad
\mathcal{A}_3(g_1^+,g_2^{+},g_3^-) = \,
\frac{{\B{1}{2}}^3}{\B{2}{3}\B{3}{1}}\ .
    \label{mhvggg}
\ee
The gluon momenta $k_i$ are assumed to be complex which
ensures  that these amplitudes do not vanish on-shell
\cite{Witten,BCFrec}.

In order to compute scattering amplitudes of gluons and massive
scalars, we need to determine the $\phi g\phi^\dag$ vertices. To
obtain these, we start with the off-shell Feynman vertex of two
scalars of mass $\mu$ and momenta $l_1,l_2$, and a single gluon
with momentum $k,$
\be V_3 (l_1^+, k^\mu, l_2^-) = \, {1 \over
\sqrt{2}}\, (l_2^\mu - l_1^{\mu}) \ .
\ee
The $\sqrt{2}$ comes
from the normalization conventions used in colour-ordered Feynman
rules \cite{Dixon}, and the $+$ and $-$ indices are labels for a
scalar and an anti-scalar. To derive the desired on-shell
amplitudes, $\mathcal{A}_3(l_1^+,k^{\pm},l_2^-)$, we contract $V_3
(l_1^+, k^\mu, l_2^-)$ with the gluon polarization vector,
$\epsilon^{\pm}_\mu (k,q)$
\be
\epsilon^{+}(k,q)_{a \dot{a}}=
\,\sqrt{2}\, {q_a\, k_{\dot{a}}
 \over  \langle q ~k\rangle} \ ,
\qquad \epsilon^{-}_\mu (k,q)= \,-\,\sqrt{2}\, {k_a \, q_{\dot{a}}
 \over [ k ~q]} \ ,
 \label{povl}
 \ee
 where $q=\ket{q}|q]$ is an arbitrary reference vector
that is not proportional to $k.$ The two independent on-shell vertices
immediately follow
\begin{eqnarray}
    \mathcal{A}_3(l_1^+,k^{+},l_2^-) &=&\, \mathcal{A}_3(l_1^-,k^{+},l_2^+)=\,
    \frac{\AA{q_1}{l_1}{k}}{\A{q_1}{k}}\ ,
    \label{threeva} \\
    \mathcal{A}_3(l_1^+,k^{-},l_2^-) &=& \, \mathcal{A}_3(l_1^-,k^{-},l_2^+)
    =\,-\frac{\AA{k}{l_1}{q_2}}{\B{q_2}{k}} \ .
    \label{threev}
\end{eqnarray}

We have already noted that the primitive vertices vanish for
on-shell real momenta in Minkowski space, but are nonzero for
on-shell complex momenta. Indeed, the on-shell conditions,
$l_1^2=l_2^2=\mu^2$ and $k^2=0,$ together with the momentum
conservation imply that the momentum of the gluon is orthogonal to
the momenta of the scalars $ k\cdot l_1=k\cdot l_2=0.$ For real
massless momentum $k$ in Minkowski space, the spinors $k^a,\tilde
k_{\dot a}$ are complex conjugates $k_{a}^\ast=\pm \tilde{k}^{\dot
a}.$ Similarly a real massive momentum forms a Hermitian matrix
$(l_{a\dot b})^\dagger=l_{b\dot a}.$ Hence for real momenta, the
conditions $l_i^{a\dot a} k_a\tilde k_{\dot a}=0,i=1,2$ imply
$l_i^{a\dot a}k_a=l_i^{a\dot a}\tilde k_{\dot a}=0$ for $i=1,2.$
It follows that the 3-point vertices (\ref{threev}-\ref{threeva})
vanish. For example we have $\CA(l_1^+,k^+,l_2^-)\propto
q_{a}l^{a\dot a}_1 \tilde k_{\dot a}=0.$

For complex momenta the spinors $k^a,\tilde k^{\dot a}$ become
independent variables. This additional freedom allows us to take
the momenta of the scalars on-shell while keeping the three-valent
amplitudes nonzero.

Finally, we note that the primitive amplitudes are
gauge-invariant, even though eqs.~\eqref{threeva}-\eqref{threev}
contain explicit $q$-dependence. Different choice of the reference
vector $q$ amounts to a gauge transformation, hence the on-shell
amplitudes should not depend on the choice of $q$ by virtue of
gauge symmetry. It is easy to see this explicitly e.g. for the
$(\phi^+ g^+ \phi^-)$ amplitude.  The reference spinor $q^a,
a=1,2$ lives in a two dimensional complex vector space. The
spinors $q^a$ and $k^a$ are independent due to the condition
$\vev{qk}\neq 0,$ so we take them as a basis of the vector space.
Hence a change in the reference spinor can be parameterized as
\qn{rech}{ q'^a=\alpha q^a+\beta k^a \ .}
Changing $q$, the
amplitude becomes
\qn{ach}{
\CA'(l_1^+,k^+,l_2^-)={\alpha\AA{q}{l_1}{k} +\beta
\AA{k}{l_1}{k}\over \alpha\vev{qk}}\ .}
Here,  $ \AA{k}{l_1}{k}=-2 k\cdot l_1=\mu^2-l_2^2=0$
is zero by momentum conservation. The
remaining $\alpha$ dependence gets cancelled between the numerator
and the denominator leaving us with the original amplitude.

It follows that the choice of the reference momenta $q_i$ of the
gluons does not affect the amplitude so in principle we could set
them to arbitrary values. In the following sections, when using
recursion relations to calculate amplitudes with scalars, we will
find it convenient to set the reference momentum of a marked gluon
in a primitive vertex \eqref{threeva} or \eqref{threev} to be the
momentum of the other marked gluon.

In the following sections we will calculate
tree-level amplitudes of the form
\be \label{trscgl2}
{\cal A}(\phi^\dagger_{l_1},g_1, g_2,
\ldots , g_m, \phi_{l_2} ) \ .
\ee
These are the colour-ordered
subamplitudes with two massive scalars and $m$ gluons $(2\le m \le 4)$ of
arbitrary helicities.

When scalars transform in the fundamental representation of the gauge group,
the `string' of fields in the amplitudes must always start and
end with the scalar, precisely as in \eqref{trscgl2}.
Using cyclic symmetry of colour-ordered amplitudes, the scalars in \eqref{trscgl2}
can be thought of as adjacent.
Scalars in the adjoint representation
can appear anywhere in the string, i.e. they do not have to be adjacent.
Such amplitudes can also be calculated straightforwardly with our methods.

We will determine all the independent helicity configurations  in \eqref{trscgl2},
all the remaining configurations can be obtained from those via the following identities:
\bea
\label{scalsym}
{\cal A}_{m+2}(\phi^{+}_{l_1},g_1^{h_1}, g_2^{h_2},\ldots, g_m^{h_m}, \phi^{-}_{l_2})
&=&
{\cal A}_{m+2}(\phi^{-}_{l_1},g_1^{h_1}, g_2^{h_2},\ldots , g_m^{h_m}, \phi^{+}_{l_2}) \\
\label{reflsym}
&=&
(-1)^m \,{\cal A}_{m+2}(\phi^{-}_{l_2},g_m^{h_m},\ldots, g_2^{h_2},g_1^{h_1}, \phi^{+}_{l_1}) \\
\label{parsym}
&=&
{\cal A}^{*}_{m+2}(\phi^{-}_{l_1},g_1^{-h_1}, g_2^{-h_2},\ldots, g_m^{-h_m},\phi^{+}_{l_2})
\eea
where $\ldots$ indicate gluon fields and $h_i$ is the helicity of the $i^{\rm th}$ gluon.
Equations \eqref{reflsym} and \eqref{parsym} follow from reflection and parity symmetry
of the colour-ordered amplitudes, and \eqref{scalsym} follows from
eqs.~\eqref{threeva}-\eqref{threev}.


\subsection{4-point amplitudes}

\begin{figure}[h]
    \psfrag{1}{$l_1^+$}
   \psfrag{u2}{$1^+$}
   \psfrag{u3}{$2^{\pm}$}
   \psfrag{2}{$\widehat{1}^+$}
   \psfrag{3}{$\widehat{2}^{\pm}$}
   \psfrag{4}{$l_2^-$}
    \psfrag{m}{\scriptsize $-$}
    \psfrag{p}{\scriptsize $+$}
    \begin{center}
        \includegraphics[width=7cm]{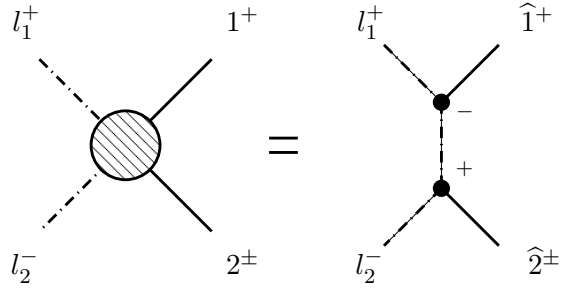}
    \end{center}
    \caption{Representation of the 4-point amplitude using the recursion relation with
    $1$ and $2$ as the shifted momenta.}
    \label{fig:4ptppp}
\end{figure}

There are two independent helicity amplitudes in this case, the
recursion relation (\ref{eq:rrel}) gives only one term for each of
the amplitudes, as illustrated in figure \ref{fig:4ptppp}.

In the case of two positive helicity gluons, 
we have:
\begin{eqnarray}
    \mathcal{A}_4(l_1^+,1^+,2^+,l_2^-) =\,
    \mathcal{A}_3(l_1^+,\widehat{1}^{+},-\widehat{P}^-)\,
    \frac{1}{P^2-\mu^2}\,
    \mathcal{A}_3(\widehat{P}^+,\widehat{2}^{+},l_2^-)\ ,
\end{eqnarray}
where we took the marked particles to be the gluons with momenta
$k_1$ and $k_2$. We set the reference vectors $q_1$ and $q_2$ of
the two gluons equal to the marked momenta gluon on the opposite
side of the diagram
\be \label{choice2}
q_2=\,\widehat{k}_1 =\,
|\widehat{k}_1 \rangle |\widehat{k}_1 ] \ , \qquad
q_1=\,\widehat{k}_2 =\, |\widehat{k}_2 \rangle |\widehat{k}_2 ]\ .
\ee
We shift the momenta along the vector $\eta=|2\rangle|1]$, so
that $|\widehat{1}] = |1] \ , \ |\widehat{2}\rangle = |2 \rangle .$
Whence, the amplitude becomes
\be
-\,\frac{\AA{2}{l_1}{1}\AA{\widehat{1}}{l_2}
    {\widehat{2}}}{\A{1}{2}^2(
    (l_1+k_1)^2-\mu^2)}\ .\nonumber
\ee
Using  $l_1+l_2+\widehat{k}_1+\widehat{k}_2 =\, 0 =\,
l_1+l_2+k_1+k_2$, this can be written as
\be
    -\,\frac{{\rm tr}(\s{\widehat{k}_1} \s{l_1} \s{\widehat{k}_2} \s{l_1})}
    {2 \A{1}{2}^2((l_1+k_1)^2-\mu^2)} \, = \,
     -\, \frac{\A{1}{2}\B{1}{2}(l_1\cdot l_1) - 2 (\widehat{k}_1\cdot l_1)
    (\widehat{k}_2\cdot l_1)
    }{\A{1}{2}^2((l_1+k_1)^2-\mu^2)}\ .\nonumber
\ee
To get the second expression we used a Fierz identity. The
second term in the numerator vanishes, $\widehat{k}_1\cdot l_1
=0$. The easiest way to see this is to use  momentum conservation
in the 3-point vertex, $l_1+\widehat{k_1}=\widehat{P}$, and the
on-shell conditions $l_1^2=\widehat{P}^2=\mu^2,$
$\widehat{k_1}^2=0.$ Alternatively this can be shown with the use
of the definition \eqref{eq:225} of $\widehat{k}$
\be
\widehat{k}_1\cdot l_1 = k_1\cdot l_1 -  \frac{(l_1+k_1)^2-\mu^2}
{2\AA{2}{l_1}{1}}\AA{2}{l_1}{1} = 0 \ . \nonumber
\ee
This leaves
us with the final answer
\begin{equation}
    \mathcal{A}_4(l_1^+,1^+,2^+,l_2^-) =\, \, \frac{ \mu^2 \B{1}{2} }
    { \A{1}{2} ((l_1+k_1)^2-\mu^2)}\ .
\end{equation}
This agrees with the previously known result computed by Bern,
Dixon and Kosower \cite{Bern:1lsdym}.

For the amplitude with one positive helicity gluon and one
negative helicity gluon the recursion relation in figure \ref{fig:4ptppp} yields,
\begin{eqnarray}
    \mathcal{A}_4(l_1^+,1^+,2^-,l_2^-) =\,
    \mathcal{A}_3(l_1^+,\widehat{1}^{+},-\widehat{P}^-)\,
    \frac{1}{\widehat{P}^2-\mu^2}\,
    \mathcal{A}_3(P^+,\widehat{2}^{-},l_2^-)\ .
\end{eqnarray}
Using the same choice for the marked gluons  and
the reference vectors \eqref{choice2} as before gives the result
\begin{eqnarray}
    \mathcal{A}_4(l_1^+,1^+,2^-,l_2^-) =\,
    -\frac{\AA{2}{l_1}{1}^2}{\A{1}{2}\B{1}{2}((l_1+k_1)^2-\mu^2)} \ ,
\end{eqnarray}
which we checked against a Feynman diagram calculation.

\subsection{5-point amplitudes}

\begin{figure}[h]
    \psfrag{l1}{$l_1^+$}
    \psfrag{u1}{$1^+$}
    \psfrag{u2}{$2$}
    \psfrag{1}{$\widehat{1}^+$}
    \psfrag{2}{$\widehat{2}$}
    \psfrag{3}{$3$}
    \psfrag{l2}{$l_2^-$}
    \psfrag{h}{\scriptsize $-$}
    \psfrag{mh}{\scriptsize $+$}
    \psfrag{mp}{\scriptsize $\mp$}
    \psfrag{pm}{\scriptsize $\pm$}
    \begin{center}
        \includegraphics[width=12cm]{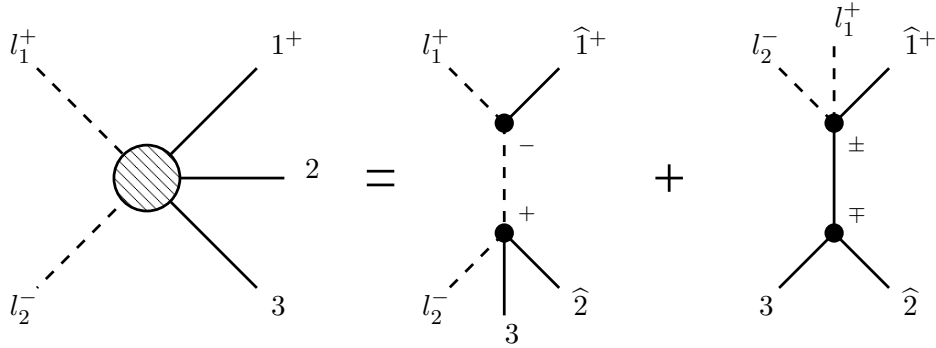}
    \end{center}
    \caption{The decomposition of the 5-point amplitude using the recursion relation with
    $1$ and $2$ as the shifted momenta.}
    \label{fig:5ptppp}
\end{figure}

The amplitudes with three gluons and a pair of scalars have three
independent helicity configurations. As before, we mark the gluons
with momenta $k_1$ and $k_2$, and pick their reference momenta to
be $q_1=\hat k_2,$ $q_2=\hat k_1.$ 
The recursion relation is depicted in figure \ref{fig:5ptppp}.
For the amplitude with all gluons
of positive helicities, the recursion relations have single
non-zero diagram. The diagram with gluon exchange vanishes as the
choice of shift vector $\eta$ implies vanishing of the $\CA(\hat
2^+,3^+,\hat p^\pm)$ \MHVb~ amplitude. The amplitude follows
immediately:
\begin{equation}
    \mathcal{A}_5(l_1^+,1^+,2^+,3^+,l_2^-) =\,  \, \frac{ \mu^2 \Bb{3}{(1+2)l_1}{1} }{
    ((l_1+k_1)^2-\mu^2)\A{1}{2}\A{2}{3}((l_2+k_3)^2-\mu^2)} \ .
    \label{eqn:314}
\end{equation}
This is in agreement with the result in \cite{Bern:1lsdym}.

For the case where one of the gluons has negative helicity we have
two independent helicity configurations, each of which has two
non-zero contributions:
\begin{eqnarray}
    \mathcal{A}_5(l_1^+,1^+,2^+,3^-,l_2^-) &=&\,
    -\, \frac{
    \AA{3}{{l}_2({1}+{2}){l}_1}{1}^2
    }
    {
    \Qa\A{1}{2}\A{2}{3}\Qc{3}\Bb{3}{({1}+{2}){l}_1}{1}
    }\nonumber\\
    &-&\frac{\mu^2\B{1}{2}^3}{s_{123}\B{2}{3}\Bb{3}{(1+2)l_1}{1}} \ ,
    \label{eqn:315}
    \\
    \mathcal{A}_5(l_1^+,1^+,2^-,3^+,l_2^-) &=&\,
    -\, \frac{ \AA{2}{l_1}{1}^2\AA{2}{l_2}{3}^2}
    {((l_1+k_1)^2-\mu^2)\A{1}{2}\A{2}{3}((l_2+k_3)^2-\mu^2)\Bb{3}{({1}+{2}){l}_1}{1}}\nonumber\\
    &+&\frac{\mu^2\B{1}{3}^4}{s_{123}\B{1}{2}\B{2}{3}\Bb{3}{(1+2)l_1}{1}} \ .
    \label{eqn:316}
\end{eqnarray}
These results are new.
Our results \eqref{eqn:314}-\eqref{eqn:316} numerically agree with the much lengthier expressions
which we obtained by a direct calculation of the 25 Feynman diagrams.

\subsection{6-point amplitudes}

\begin{figure}[h]
    \psfrag{l1}{$l_1^+$}
    \psfrag{u1}{$1^+$}
    \psfrag{u2}{$2$}
    \psfrag{1}{$\widehat{1}^+$}
    \psfrag{2}{$\widehat{2}$}
    \psfrag{3}{$3$}
    \psfrag{4}{$4$}
    \psfrag{l2}{$l_2^-$}
    \psfrag{h}{\scriptsize $-$}
    \psfrag{mh}{\scriptsize $+$}
    \psfrag{mp}{\scriptsize $\mp$}
    \psfrag{pm}{\scriptsize $\pm$}
    \begin{center}
        \includegraphics[width=14cm]{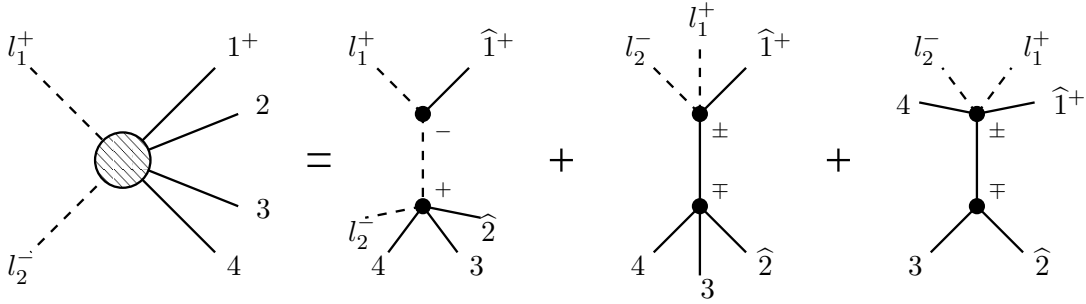}
    \end{center}
    \caption{The decomposition of the 6 point amplitude using the recursion relation
    with $1$ and $2$ as the shifted momenta.}
    \label{fig:6pt}
\end{figure}

We mark gluon momenta 1 and 2, and write down the recursion
relation for the 6-point amplitudes with 4 gluons in figure
\ref{fig:6pt}.

In the case of all gluons of the same helicity, only the first
diagram contributes. We find,
\begin{equation}
\label{6ptplus}
    \mathcal{A}_6(l_1^+,1^+,2^+,3^+,4^+,l_2^-) =\,
    -\,\frac{ \mu^2 \Bb{4}{l_2(3+4)(1+2)l_1}{1}
    }{Q_1Q_2Q_3\A{1}{2}\A{2}{3}\A{3}{4}}\ ,  \nonumber
\end{equation}
where $Q_1=\Qa$, $Q_2=\Qb$ and $Q_3=\Qc{4}$.
Eq.\ref{6ptplus} is a slightly shorter form of the result given in~\cite{Bern:1lsdym}.

Now we compute the remaining independent 6-point
amplitudes.  There are two amplitudes with one negative helicity gluon:
\begin{align}
    &\mathcal{A}_6(l_1^+,1^+,2^+,3^+,4^-,l_2^-) = \nonumber\\
    &
    +\frac{
    \left (
    Q_2\AA{4}{{l}_2({1}+{2}+{3}){l}_1}{1}
    -\mu^2(\AA{4}{{l}_2{3}{2}}{1})
    \right )^2}
    {
    Q_1 Q_2 Q_3 \A{1}{2}\A{2}{3}\A{3}{4}
    \Bb{4}{l_2(3+4)(1+2)l_1}{1}
    } \nonumber \\
    &
    +\frac{
    \mu^2\Bb{3}{({1}+{2}){l}_1}{1}^3
    }
    {
    Q_1\A{1}{2}\B{3}{4}
    \AA{2}{({3}+{4})({l}_1+{l}_2){l}_1}{1}
    \Bb{4}{l_2(3+4)(1+2)l_1}{1}
    } \nonumber\\
    &
    -\frac{
    \mu^2\AA{4}{{2}+{3}}{1}^3
    }
    {
    s_{1234}s_{234}\A{2}{3}\A{3}{4}\AA{2}{(3+4)(l_1+l_2)l_1}{1}
    }
    \label{eq:6ptpppm}
    \hspace{0pt}\intertext{.}
    &\mathcal{A}_6(l_1^+,1^+,2^+,3^-,4^+,l_2^-) = \nonumber\\
    &
    +\frac{
    (Q_2\AA{3}{{l}_1}{1}-\mu^2\B{3}{2}\B{2}{1})^2\AA{3}{{l}_2}{4}^2
    }
    {
    Q_1  Q_2 Q_3 \A{1}{2}\A{2}{3}\A{3}{4}
    \Bb{4}{l_2(3+4)(1+2)l_1}{1}
    } \nonumber \\
    &
    +\frac{
    \mu^2\Bb{4}{({1}+{2}){l}_1}{1}^4
    }
    {
    Q_1\A{1}{2}\B{3}{4}\Bb{3}{({1}+{2}){l}_1}{1}
    \AA{2}{({3}+{4})({l}_1+{l}_2){l}_1}{1}
    \Bb{4}{l_2(3+4)(1+2)l_1}{1}
    } \nonumber\\
    &
    -\frac{
    \mu^2\AA{3}{{2}+{4}}{1}^4
    }
    {
    s_{1234}s_{234}\AA{4}{{2}+{3}}{1}\A{2}{3}\A{3}{4}\AA{2}{(3+4)(l_1+l_2)l_1}{1}
    } \nonumber \\
    &
    +\frac{
    \mu^2\Bb{4}{({1}+{2}+{3}){l}_1}{1}\B{1}{2}^3
    }
    {
    s_{123}\AA{4}{{2}+{3}}{1}Q_3\B{2}{3}\Bb{3}{(1+2)l_1}{1}
    }
    \label{eq:6ptppmp}
\end{align}
These amplitudes agree with the massless MHV-type amplitudes as $\mu^2\to 0$.
There are three independent helicity amplitudes with two negative helicity gluons:
\rm
\begin{align}
    &\mathcal{A}_6(l_1^+,1^+,2^+,3^-,4^-,l_2^-) = \nonumber\\
    &+\frac{
    \left (
    Q_2\AA{4}{{l}_2({3}+{4}){l}_1}{1}-\mu^2\AA{4}{{l}_2({3}+{4}){2}}{1}
    \right )^2
    }
    {
    Q_1Q_2Q_3\A{1}{2}\B{3}{4}\Aa{4}{l_2({3}+{4})}{2}\Bb{3}{({1}+{2}){l}_1}{1}
    } \nonumber \\
    &+\frac{
    \mu^2\AA{2}{{l}_1}{1}^2\A{3}{4}^3
    }
    {
    Q_1\A{1}{2}\A{2}{3}\AA{2}{({3}+{4})({l}_1+{l}_2){l}_1}{1}\Aa{4}{l_2({3}+{4})}{2}
    } \nonumber \\
    &+\frac{
    \Bb{1}{l_2l_1}{1}^2\A{3}{4}^3
    }
    {
    s_{1234}s_{234}\AA{4}{{2}+{3}}{1}\A{2}{3}\AA{2}{(3+4)(l_1+l_2)l_1}{1}
    } \nonumber \\
    &-\frac{
    \AA{4}{{l}_2({1}+{2}+{3}){l}_1}{1}^2\B{1}{2}^3
    }
    {
    Q_3s_{123}\AA{4}{{2}+{3}}{1}\Bb{4}{({1}+{2}+{3}){l}_1}{1}\B{2}{3}\Bb{3}{(1+2)l_1}{1}
    } \nonumber \\
    &+\frac{
    \mu^2\B{1}{2}^3
    }
    {
    s_{1234}\B{2}{3}\B{3}{4}\Bb{4}{(1+2+3)l_1}{1}
    }
\label{eq:6ptppmm}
\hspace{0pt}\intertext{}
    &\mathcal{A}_6(l_1^+,1^+,2^-,3^+,4^-,l_2^-) = \nonumber\\
    &-\frac{
    \AA{2}{l_1}{1}^2\AA{2}{l_1+1}{3}^2\AA{4}{l_2}{3}^2
    }
    {
    Q_1Q_2Q_3\A{1}{2}\B{3}{4}\Aa{4}{l_2(3+4)}{2}\Bb{3}{(1+2)l_1}{1}
    } \nonumber \\
    &-\frac{
    \mu^2\A{2}{4}^4\AA{2}{{l}_1}{1}^2
    }
    {
    Q_1\A{1}{2}\A{2}{3}\A{3}{4}\AA{2}{({3}+{4})({l}_1+{l}_2){l}_1}{1}\Aa{4}{l_2(3+4)}{2}
    } \nonumber \\
    &+\frac{
    \Bb{1}{l_2l_1}{1}^2\A{2}{4}^4
    }
    {
    s_{1234}s_{234}\AA{4}{2+3}{1}\A{2}{3}\A{3}{4}\AA{2}{(3+4)(l_1+l_2)l_1}{1}
    } \nonumber \\
    &-\frac{
    \AA{4}{l_2(1+2+3)l_1}{1}^2\B{1}{3}^4
    }
    {
    Q_3s_{123}\B{1}{2}\B{2}{3}\Bb{4}{(1+2+3){l}_1}{1}\Bb{3}{(1+2)l_1}{1}\AA{4}{2+3}{1}
    } \nonumber \\
    &+\frac{
    \mu^2\B{1}{3}^4
    }
    {
    s_{1234}\B{1}{2}\B{2}{3}\B{3}{4}\Bb{4}{(1+2+3)l_1}{1}
    }
    \label{eq:6ptpmpm}
\end{align}

\begin{align}
    &\mathcal{A}_6(l_1^+,1^+,2^-,3^-,4^+,l_2^-) = \nonumber\\
-&\frac{\AA{2}{l_1}{1}^2(Q_3\BB{4}{l_1+1}{2}-\mu^2\B{4}{3}\A{3}{2})^2}
{Q_1Q_2Q_3 \A{1}{2}\B{3}{4}\Bb{3}{(1+2)l_1}{1}\Aa{4}{l_2(3+4)}{2}}\nonumber \\
-&\frac{\mu^2\A{2}{3}^3 \AA{2}{l_1}{1}^2}{Q_1 \A{1}{2}\A{3}{4}
\Aa{4}{l_2(3+4)}{2}
\AA{2}{(3+4)(l_1+l_2)l_1}{1}}\nonumber \\
+&\frac{\Bb{1}{l_2l_1}{1}^2\A{2}{3}^3}{s_{1234}s_{234} \A{3}{4}
\AA{2}{(3+4)(l_1+l_2)l_1}{1} \AA{4}{2+3}{1}}\nonumber \\
-&\frac{\Bb{1}{(2+3)l_1}{1}^2\Bb{1}{(2+3)l_2}{4}^2}
{s_{123} Q_3 \B{1}{2}\B{2}{3} \Bb{3}{(1+2)l_1}{1}\Bb{4}{(1+2+3)l_1}{1}
\AA{4}{2+3}{1}}\nonumber \\
+&\frac{\mu^2\B{1}{4}^4}{s_{1234}\B{1}{2}\B{2}{3}\B{3}{4} \Bb{4}{(1+2+3)l_1}{1}}
    \label{eq:6ptpmmp}
\end{align}
All of the above six-point amplitudes have been checked numerically against an 
independent calculation of the 220 Feynman diagrams.



\section{Vanishing of  $A(z)$ at Infinity}
\label{vanish}

The recursion relations are valid as long as ${\cal
A}(z)\rightarrow 0$ for $z\rightarrow \infty.$ This is the case
only for some choices of marked particles.  Indeed, for the marked
gluons $i,j$ with
\qn{shifts}
{\hat{|i\rangle}=|i\rangle+z|j\rangle \ , \qquad
\hat{|j]}=|j]-z|j]\ ,}
it was observed in \cite{BCFrec}  that
$\CA(z)\rightarrow 0$ for the helicity assignments of gluons
$(h_i,h_j)=(+,+),(+,-),(-,-).$ This has been extended to fermions
in \cite{Luo:2005rx,Luo:2005my}. A direct diagrammatic proof for
gluon amplitudes in the $(+,-)$ case was given in
\cite{Britto:2005fq}. Here, we present an argument valid for all
three helicity assignments of marked gluons putting no
restrictions on the remaining partons\footnote{For the $(+,+)$
case, our proof works for the amplitudes that have two extra
negative helicity gluons. However, see the end of this section for
a heuristic argument, which removes this assumption.}. We also
consider the case where one or both of the marked particles are
massless fermions.

\subsection{The $(+,-)$ case}

We start with the $(+,-)$ case for which we will show that all
Feynman diagrams contributing to ${\cal A}(z)$ vanish at infinity.
In a given Feynman diagram, the $z$-dependence flows along a
unique path of Feynman propagators and vertices.  The Feynman
diagrams with most dangerous $z$-dependence are those in which all
vertices along this path are trivalent. These vertices are linear
in momentum so they each contribute a factor of $z.$ Each of the
propagators gives a factor of $1/z.$ A path made of $r$
propagators has $r+1$ vertices. Hence, the propagators and
vertices go at infinity as $z^{r+1}/z^r=z.$ For fermions, the
propagator $(\s{p}+m)/(p^2-m^2)$ goes like a constant at infinity,
but so do the vertices containing fermions.  Each of the fermion
propagators increases the above estimate by a factor of $z$ and
each of the fermion vertices decreases the estimate by $z.$ For
$f$ fermion propagators, there are at least $f+1$ fermion vertices
along the path of the $z$-dependence, hence the above bound on the
Feynman diagram gets improved by at least a factor of $1/z$.

The remaining pieces of the Feynman diagram that  depend on $z$
are the polarization vectors of the marked gluons.
Consider marked gluons of opposite helicity, so that we can have either
 $g_i^+,$ $g_j^-$, or $g_i^-,$ $g_j^+$.
The spinors $\l_i (z)$ and $\tilde\l_j (z)$ are linear in $z$,
\eqn{lzet}, while $\tilde\l_i(z),\l_j(z)$ are independent of $z.$
It follows from \eqn{povl} that for the helicity configuration
$g_i^+,g_j^-$, the polarization vectors of both gluons  go as
$1/z$ at infinity. Altogether Feynman diagrams vanish at infinity
as $\CO(z/z^2)=\CO(1/z).$

In the opposite case, $g_i^-,$ $g_j^+$, gluon polarization vectors
grow as $z$ at infinity and individual Feynman diagrams go as
$\CO(z^3)$ leading to nontrivial boundary contributions to the
recursion relations for ${\cal A}(z)$.

If one of the marked particles is a gluon and the other a massless
fermion, we can proceed analogously. Now the propagators and
vertices contribute a factor of at worst $z^0$ because of the fermions. For
helicity assignments $(h_i,h_j)=(+1,-1/2),(+1/2,-1),$ the gluon
polarization vector goes like $1/z$ and the fermion polarization
spinors $u^-(\widehat{p}_j)=\lambda_j^a$ and
$u^+(\widehat{p}_i)=\tl_i^{\dot a}$ are independent of $z.$ Hence
the shifted amplitude vanishes at infinity for the assignment
$(h_i,h_j)=(+,-).$

\subsection{The $(+,+)$ and $(-,-)$ cases}

The $(h_i,h_j)=(+,+)$ case cannot be treated by counting powers of
$z$ in {\it individual} Feynman diagrams and requires a more
elaborate argument along the lines of \cite{Cachazo:2005ca}. The
argument works for amplitudes that, besides gluons $g_i^+,g_j^+$
have two extra negative helicity gluons $g_k^-,g_l^-,$ hence in
particular it works for all QCD gluon amplitudes since at tree
level, the gluon amplitudes always have at least two gluons of
both helicities.

Consider the function ${\cal A}(\lambda_m(z),\tl_m(z))$
constructed from the scattering amplitude $\CA$ by shifting the
spinors
\qalign{zdil}
{\ket{i}&=&\ket{i}+z\ket{k}+z\ket{l}\cr
|k]&=&|k]-z|i]\cr |l]&=&|l]-z|i] }
while keeping $|i],$
$\ket{k}$ and $\ket{l}$ unshifted. The shifts \eqref{zdil}
maintain momentum conservation, so $\CA(z)$ can be computed from
usual Feynman diagrams.

\begin{figure}[h]
 \psfrag{i}{$i^+$}
 \psfrag{j}{$k^-$}
 \psfrag{k}{$l^-$}
 \centering
 \includegraphics[height=1.5in]{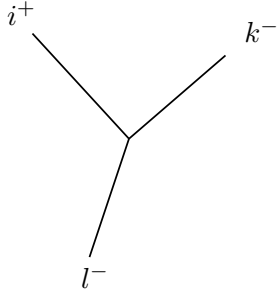}
 \caption{ In a Feynman diagram with marked  particles $i,k,l$,
 the $z$ dependence flows along a unique three-legged
 path of propagators and vertices.}
 \label{threel}
\end{figure}

Note that the function $\CA(z)$ vanishes
at infinity. As in the previous case, we prove this by studying
the most dangerous Feynman diagrams. The $z$-dependence flows
along a `three-legged path' of Feynman propagators and vertices,
with each leg ending at one of the external gluons $i,k,l$
This path is illustrated in figure \ref{threel}.
For a path consisting of $r$ propagators, there
are $r+1$ vertices. For large $z,$ the propagators and vertices
give a $z^{r+1}/z^r=z$ contribution to the $z$-dependence of
$\CA(z).$ The polarization vectors of each of the gluons give a
factor of $1/z$ so the function $\CA(z)$ vanishes for large $z$ as
$\CO(1/z^2).$

Hence, following the discussion of recursion relations in section
{\bf \ref{sec:mrc}}, we have
\qn{relv} {\CA=\sum_{\alpha,h}  {\cal
A}_L(z_\alpha,-P_\alpha(z_\alpha)^h) {1\over
P_\alpha^2-m_\alpha^2}{\cal
A}_R(z_\alpha,P_\alpha(z_\alpha)^{-h})\ ,}
where the summation is
over partitions of the particles into two sets $L,R$ such that
$i\in L$ and at least one of $j,k$ is in $R$, and also over
helicities of the intermediate state $P_\alpha .$

The amplitudes
$\CA_L,\CA_R$ are evaluated at
\qn{zalpha}
{z_\alpha = \, {P_\alpha^2-m_\alpha^2\over
\AA{k}{P_\alpha}{i}+\AA{l}{P_\alpha}{i}}\ .}
Similar recursion
relations with three or more marked particles have been recently
considered in \cite{Bern:2005hs, Bedford:2005yy, Cachazo:2005ca}.

We will use this expression for $\CA$ in an induction proof of the
recursion relations with  $(h_i,h_j)=(+,+).$ We consider the
function $\CA(y)$ with
\qn{ijshift}
{\ket{i}=\ket{i}+y\ket{j}\qquad |j]=|j]-y|i]}
and estimate the $y$
dependence of $\CA (y)$ with the help of \eqn{relv}. For a
partition $\alpha$ such that  $i,j$ are both in $L,$ neither
$P_\alpha$ nor $z_\alpha$ depend on $y.$ So the entire $y$
dependence comes from $\CA_L.$ But $\CA_L$ has fewer particles
than $\CA$ so by induction hypothesis it vanishes as $y$ goes to
infinity.

The other case with $i\in L, j\in R$ is more subtle. We have
$P_\alpha(y)=P_\alpha + y\l_j\tl_i$ so $z_l$ depends linearly on
$y$
\qn{liny} {z_\alpha=\,
{P_\alpha^2-y\AA{j}{P_\alpha}{i}-m_\alpha^2\over\
\AA{k}{P_\alpha}{i}+\AA{l}{P_\alpha}{i}}\ .}
We represent the
on-shell amplitudes $\CA_L,\CA_R$ as a sum of Feynman diagrams.
The $y$ dependence flows along a four-legged tree with each leg of
the tree ending at one of the gluons $i,j,k,l.$ A tree composed of
$r$ propagators (one of which is the propagator connecting $\CA_L$
and $\CA_R$), contains $r+1$ vertices. The most dangerous
contribution comes from Feynman diagrams in which the $r+1$
vertices are three-point vertices that each give a factor of $y.$
Hence, the propagators give a factor of $1/y^r$ and the vertices a
factor of $y^{r+1}.$

The polarization vectors of the gluons $i,k,l$ each give a factor
of $1/y$ and that of the gluon $j$ a factor of $y,$ so altogether
we have $1/y^2$ coming from the polarization vectors. There are
two more `external' particles with $y$ dependent momentum. These
come from the ends of the internal line connecting $\CA_L$ and
$\CA_R.$ They contribute  $\sum_h \epsilon^h\epsilon^{-h}$ which,
in a convenient gauge, goes like a constant at infinity. For
example for a gluon we have
\qn{gpol}
{\sum_{h=\pm}\epsilon^h_\mu\epsilon^{-h}_\nu=g_{\mu\nu}-
{P_{\mu}q_\nu+P_{\nu}q_\mu\over P\cdot q}\ ,}
where $P=P_\alpha(y)$
and $q$ is the reference momentum of the internal gluon. For large
$y,$ this is independent of $y.$ Hence, $\CA(y)$ vanishes as
$\CO(y^{r+1}/(y^ry^2))=\CO(1/y)$ for large $y.$

In this proof, we have assumed that the amplitude has besides the
two marked positive helicity gluons $i,j$ also two negative
helicity gluons $k,l$. However we believe that this restriction is
not necessary. Heuristically, one can argue as follows. Consider
any amplitude $\CA$ without imposing the restriction that it
contains two negative helicity gluons. We can construct an
`auxiliary' amplitude ${\cal A}'$ which, besides external
particles present in $\CA$ also has two extra negative helicity
gluons $k,l.$ In the previous argument, we demonstrated that
$\CA'(y)$ with marked gluons $i,j$ vanishes for large $y$ as
$\CO(1/y).$ One can imagine taking soft gluon limit
$p_l,p_k\rightarrow 0$ to recover the original amplitude. Taking
the soft limit gives back a universal factor, times an amplitude
with the gluon removed. The universal factor depends on the
momentum of the soft gluon and of the possible adjacent particles.
Hence, as long as the `auxiliary' gluons $k,l$ are not adjacent to
the marked gluons $i,j,$ their soft factors are $y$ independent.
So the $y$ asymptotic of the amplitude $\CA$ with gluons $k,l$
removed should be the same as that of $\CA'.$

\section{Summary}

In this paper we have constructed tree-level recursion relations valid in
any quantum field theory which naturally incorporate massive particles. 
As an application, we focus on scalar particles.   However, there is no
obstacle in applying our general method to massive particles with spin.
With the general recursion relations in place, one may be able to avoid
in future all Feynman-diagrams calculations of nontrivial tree
amplitudes.

As a first application of these recursion relations, we have derived 
expressions for scattering amplitudes involving a pair of massive scalars
and up to four gluons. Explicit results are given in Section~{\bf 3}.
These compact analytic expressions for amplitudes may be useful in
deriving the one-loop six gluon amplitudes in QCD using the unitarity
method of Bern et al~\cite{Bern:1995db,Bern:1lsdym}.

\vskip .3in
\begin{center}
{\bf Acknowledgements} \end {center}

We would like to thank F. Cachazo, G. Travaglini and E. Witten for
useful discussions. EWNG and VVK acknowledge the support of PPARC
through Senior Fellowships and SDB acknowledges the award of a
PPARC studentship. The work of P. Svr\v{c}ek was supported in part
by Princeton University Centennial fellowship and by NSF grants
PHY-9802484 and PHY-0243680. Any opinions, findings and
conclusions or recommendations expressed in this material are
those of the authors and do not necessarily reflect the views of
the funding agencies

\bibliographystyle{JHEP-2}

\bibliography{refs}

\end{document}